\documentclass[a4paper,11pt]{article}
\usepackage{pos}

\title{The AdvCam project: Designing the future cameras for the Large-Sized Telescope of the Cherenkov Telescope Array Observatory}
\ShortTitle{The AdvCam project}

\author*[a]{M. Heller}
\onbehalf{on behalf of the CTAO-LST Project}

\affiliation[a]{D\'epartement de Physique Nucl\'eaire, Facult\'e de Sciences, Universit\'e de Gen\`eve, \\
Quai Ernest Ansermet 24, CH-1205 Gen\`eve, Switzerland}

\emailAdd{matthieu.heller@unige.ch}

\abstract{
An international collaboration composed of Italian, Japanese, Spanish and Swiss institutes, is developing the advanced camera (AdvCam), the next-generation camera for Imaging Atmospheric Cherenkov Telescopes, designed specifically for the Large-Sized Telescopes (LST) of the Cherenkov Telescope Array Observatory.
AdvCam incorporates cutting-edge Silicon Photomultipliers (SiPMs) and a fully digital readout system, setting new standards for performance and efficiency.

The upgraded camera will feature four times more pixels for the same field of view as the existing PMT-based camera, enabling finer image resolution and significantly improving angular precision and background noise rejection. 
To cope with the increase in number of pixels, many technological challenges are being tackled, from low power and high speed integrated chip design to real-time data processing on hardware accelerators.

This technological leap will lower the energy threshold by allowing operation at lower observation threshold and providing brighter images. 
The increase in effective area, angular and energy resolution will enhance the sensitivity, unlocking new potential for gamma-ray astronomy. 
In this work, we present the performance of the AdvCam’s core building blocks and its innovative architecture capable of enabling unprecedented triggering capabilities. We also showcase the latest performance results based on Monte-Carlo data that has been tuned to reflect the latest stages of the on-going technological developments, highlighting the transformative capabilities of this next-generation instrument.
}

\FullConference{39th International Cosmic Ray Conference (ICRC 2025) in Geneva, Switzerland}

\begin{document}
\maketitle

\section{Introduction}

The Advanced Camera (AdvCam)~\cite{Heller2023} is a next-generation instrument proposed for the long-term operation of the Large-Sized Telescopes (LSTs)~\cite{Mazin2017} of the Cherenkov Telescope Array Observatory (CTAO), which is expected to operate for $\sim$30 years. 
This exceeds the typical lifespan of current photomultiplier tube (PMT)-based cameras, making a robust, low-maintenance, and high-performance alternative increasingly attractive. AdvCam is designed to offer improved sensitivity, extended sensor durability and intelligent data processing to suppress background more efficiently close to the detector level.

The system builds on the expertise gained from the current LST and SST-1M SiPM camera developments~\cite{Alispach2025}, integrating fully digitizing electronics, reprogrammable trigger logic, and AI-based trigger and data handling. 
In this work we present the camera design, the current status of its implementation and its performance derived on Monte Carlo simulations.

\section{The AdvCam design}

\begin{figure}
    \centering
    \includegraphics[width=0.29\linewidth]{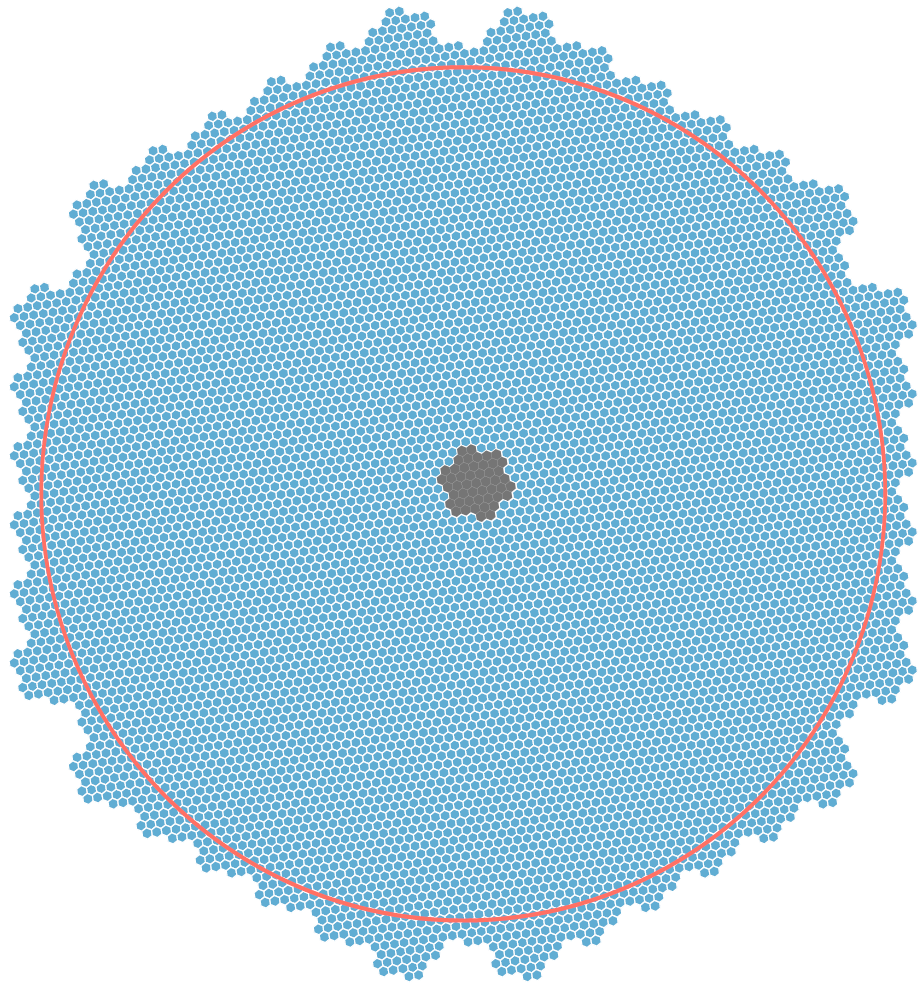} \hfill
    \includegraphics[width=0.70\linewidth]{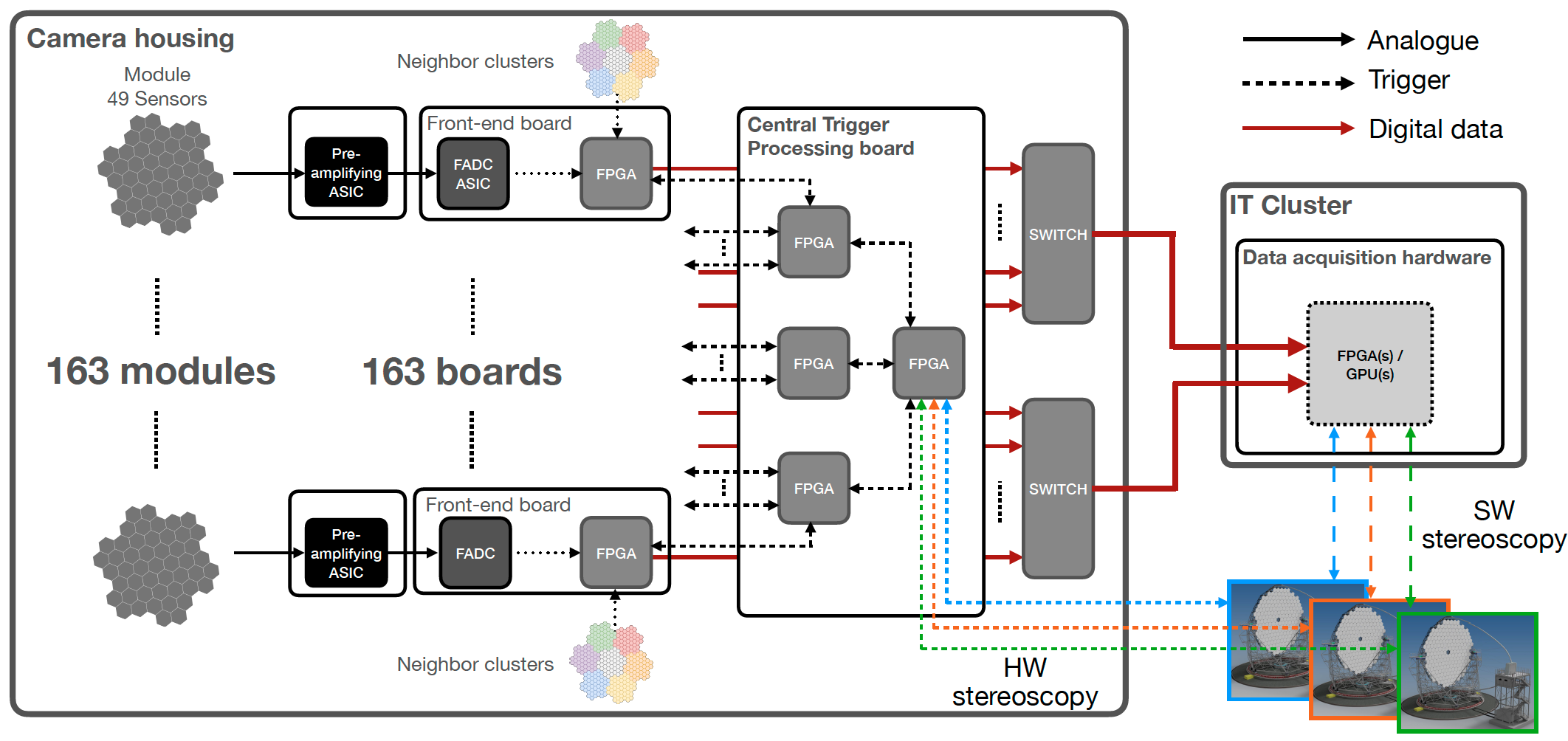}
    \caption{Left: View of the AdvCam geometry, built out of 7987 pixels. The greyed region in the center shows the size of one optical module while the red circle illustrate a field of view of 4.3$^{\circ}$, equal to the existing camera one. Right: Simplified readout architecture, from the pre-amplification of the SiPM signals to the storage in the IT cluster of CTAO.}
    \label{fig:advcam}
\end{figure}

\subsection{The overall architecture}
The proposed camera, which geometry is shown in Figure~\ref{fig:advcam}--left, will be composed of 163 modules of 49 hexagonal pixels each for a total of 7987 pixels and a minimum field of view of 4.3$^{\circ}$. Each pixel features a hollow light guide with an opening of linear size $\sim$2.4~cm corresponding to $\sim$0.05$^{\circ}$. A hexagonal SiPM of latest generation is located at the output. The pixels are arranged in group of seven, a central one and six neighbours, called flower. Seven flowers are put together to form a super-flower of 49 pixels.
As shown in Figure~\ref{fig:advcam}--right, the SiPM output is amplified by an application specific integrated circuit (ASIC) whose main design drivers are the speed and the power consumption. The analogue output of the ASIC is fed to a fast analogue to digital converter (FADC) whose digital output is streamed to powerful field programmable gate array (FPGA), common to the 49 pixels of a module, located on the front-end board (FEB). 
Each FEB also connects to the FEBs of the six neighbouring modules in order to build the L1 trigger which is described in \ref{sec:l1trig}. 
When a positive L1 trigger decision is registered by a module, a signal is sent to the Central Trigger Processor Board (CTPB) which can perform a more advanced trigger decision based on the information collected by all FEBs.

As detailed in \cite{Burmistrov2025}, spatio-temporal clustering as well as real-time inference using artificial intelligence algorithms can be deployed on the CTPB for further data volume reduction. The CTPB also connects to all other surrounding large-sized telescopes to build a hardware stereo trigger. 

If the event is validated, the CTPB will issue a trigger signal to all front-end boards, prompting them to transmit data via the Remote Direct Memory Access (RDMA) over Converged Ethernet (RoCE) protocol to compatible network and data processing hardware located in the observatory building where the data acquisition and on-site data analysis will run. This protocol was chosen to enable the use of commercially available off-the-shelf components.

\subsection{Photo detection plane}
The development of the photo detection plane is performed jointly by the UniGe together Nagoya University and the Institute of Cosmic Ray Research.
The photo detection plane will be composed of hexagonal SiPMs, identical in size to those used in \cite{Burmistrov2025}. The baseline technology for the sensor is the S13360-3050/75CN-UVE, which is a evolution of the S13360 technology with enhanced sensitivity to UV and faster response time. As shown in \cite{Abe2023}, to achieve very short single photoelectron response, the first prototype sensor features a large quenching resistor which affects the stability of the SiPM response under varying Night-Sky Background (NSB) conditions. 
This resistor will be decreased in the final sensor version to maintain the pulse full width half maximum while decreasing by four the recharge time. Since the first development, Hamamatsu has also developed a process to apply optical coating onto the SiPM surface to cut-out photons with wavelengths $>$540~nm thus rejecting the first peak of the background light spectrum. 
Alternatively, as shown in \cite{Abe2023}, the coating of the light-guides can be adapted to perform this task although with a lower efficiency as a non-negligible fraction of the light reach the sensor surface without being reflected on the light-guide surface.

\subsubsection{The pre-amplifying ASIC}
\label{sec:preamp}
The front-end electronics of the telescope camera play a critical role in ensuring the accurate and efficient readout of signals from the silicon photomultipliers (SiPMs). The main design challenges consist in preserving a fast response and single photoelectron resolution while managing the high capacitance associated with large-area sensors and minimizing the power consumption. 
In order to verify the compliance with the application requirements, a prototype ASIC was developed at University of Geneva (UniGe) using a standard CMOS 65~nm technology node.
The ASIC offers several key features: AC coupling with the camera pixels to avoid baseline shifts caused by NSB; active summation for optimal signal reconstruction with improved signal-to-noise ratio; support for multiple digitization schemes via dedicated pseudo-differential and single-ended processing channels; and compatibility with various SiPM types through tunable capacitors and amplifier current settings.

\begin{figure}
    \centering
    \includegraphics[width=0.54\textwidth]{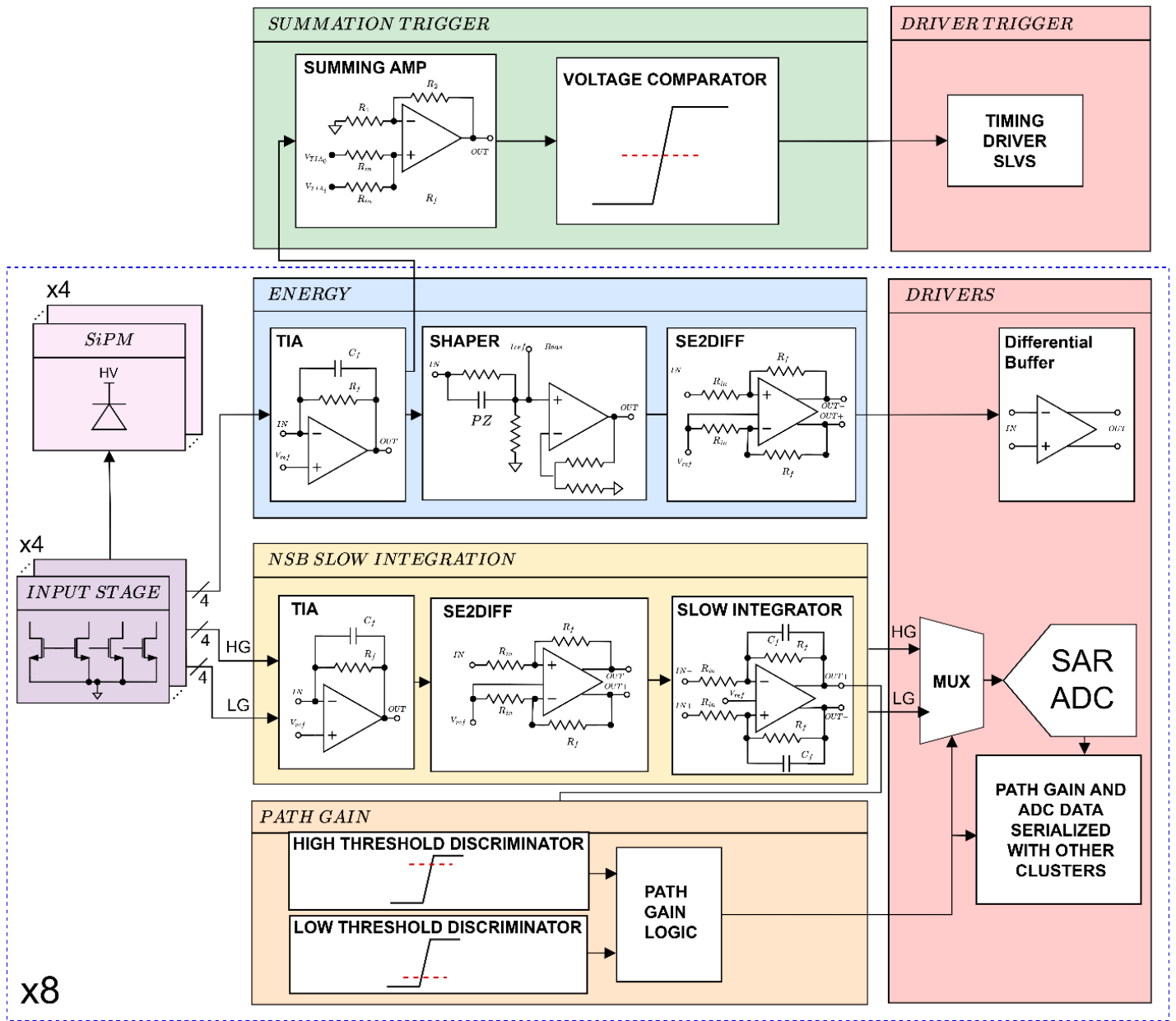}
    \includegraphics[width=0.44\textwidth, trim={1cm, 0cm, 0.5cm, 0cm}, clip]{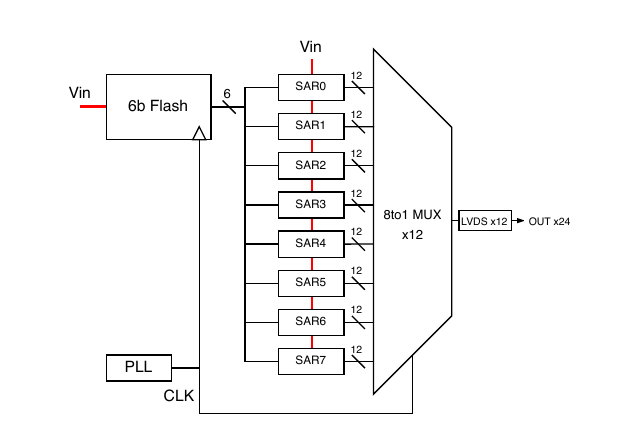}
    \caption{Left: Architecture of the PRESSEC ASIC being developed for the AdvCam. Right: Block diagram of the FADC ASIC}
    \label{fig:fansic_schem}
\end{figure} 

The ASIC has been tested together with the S13360-3075CS-UVE SiPM from Hamamatsu. Using a high-gain configuration for the ASIC provides an optimal noise performance which allows to distinguish individual peaks up to $\sim30\,$ photo-electrons (p.e.). The signal-to-noise ratio\footnote{defined as $\mathit{SNR}(N_\mathrm{pe}) = G(N_\mathrm{pe})/\sqrt{\sigma_e^2 + N_\mathrm{pe}\,
\sigma_s^2}$, where $G(N_\mathrm{pe})$ is the electronics gain at at given number of p.e. $N_\mathrm{pe}$, $\sigma_e$ and $\sigma_s$ being the electronics and sensor noise.} (SNR) is reaches its maximum at 1\,p.e., of $6.98\pm0.22$. 
The ASIC's dynamic range, with high gain, extends up to 800\,p.e. with a power consumption of 23~mW per pixel. Moreover, within the expected operational range of 1--250\,p.e., FANSIC provides a full width at half maximum (FWHM) of 3\,ns and a post-calibration linearity within 5\%. The full characterization results are detailed in~\cite{Giangrande2025}. 
A second iteration of the ASIC has been designed and will be submitted during the summer of 2025, featuring eight single ended and differential outputs with improved pulse shape.

The final ASIC, the PRESSEC (Preamplifier Readout Electronics for Summing SiPMs Enhanced Circuit) developed in collaboration with University of Barcelona and Polytechnic University of Catalonia, will have a different architecture with enhanced performance and functionalities as shown in Figure~\ref{fig:fansic_schem}. The energy path, which will provide continuous amplified waveform will have the same characteristics as the FANSIC ASIC in term of SNR, dynamic range and speed. The linearity should be kept below 3\% without calibration. 
A slow integrator with adaptable gain will allow monitoring of the night sky background level useful for implementing feedback loop for the trigger and SiPM signal conditioning. The analogue to digital converted for this path will added to the ASIC. Not shown in this scheme, it will be possible to inject a calibration pulse to test the sanity of the readout chain. Eventually, and even though not necessary as the trigger will be digital, an analogue trigger path will be also available.

\subsubsection{The FADC ASIC}
Two fast ADC ASIC prototypes for the AdvCam have been developed by the group of EPFL/AQUA\footnote{École Polytechnique Féderale de Lausanne, Advanced Quantum Architecture Laboratory}. Both designs aim to achieve $>$800~MS/s sampling, 12-bit resolution, and $<$200~mW power. The first version of the ASIC, fabricated in 110~nm, combined Flash and time-interleaved successive approximation register (TI-SAR) stages but failed to meet specifications due to architectural and process limitations, yielding only 3-bit output.
The second version of the ASIC, implemented in 65~nm CMOS, uses a 6-bit Flash front-end followed by 8x6-bit TI-SAR ADCs and achieves 1~GS/s with 12-bit resolution (see Figure~\ref{fig:fansic_schem}). 
Key system components—phase locked loop, low voltage differential signaling and serial peripheral interface—are operational, and measurement campaigns are ongoing. Early results from the 6-bit Flash indicate correct high-speed functionality (800~MHz), with TI-SAR testing in progress. The full chain, including FPGA readout and high-speed serialization, is in place for integration with preamplifiers and system-level testing.

\subsubsection{The Front-End board}
\begin{figure}
    \centering
    \includegraphics[width=0.48\textwidth]{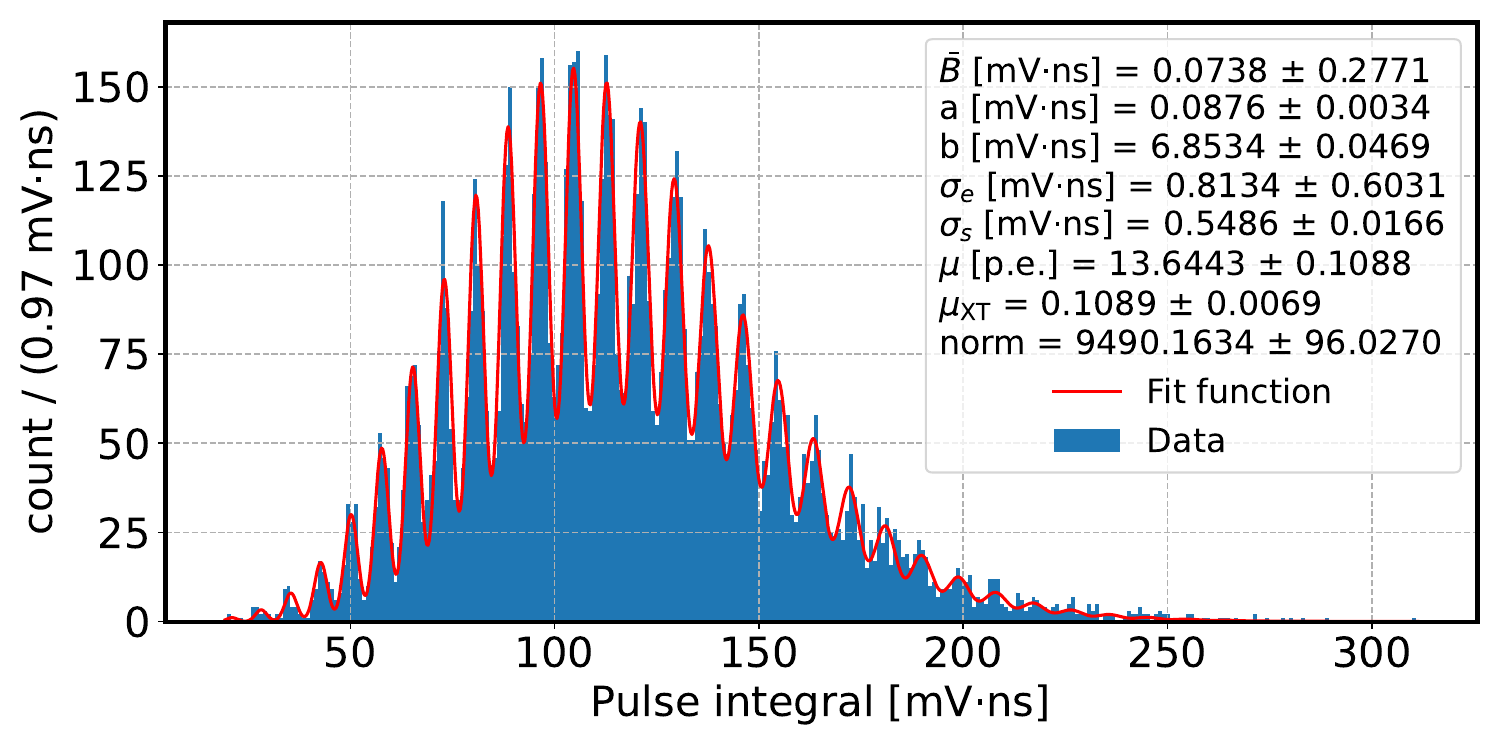}
    \includegraphics[width=0.48\textwidth]{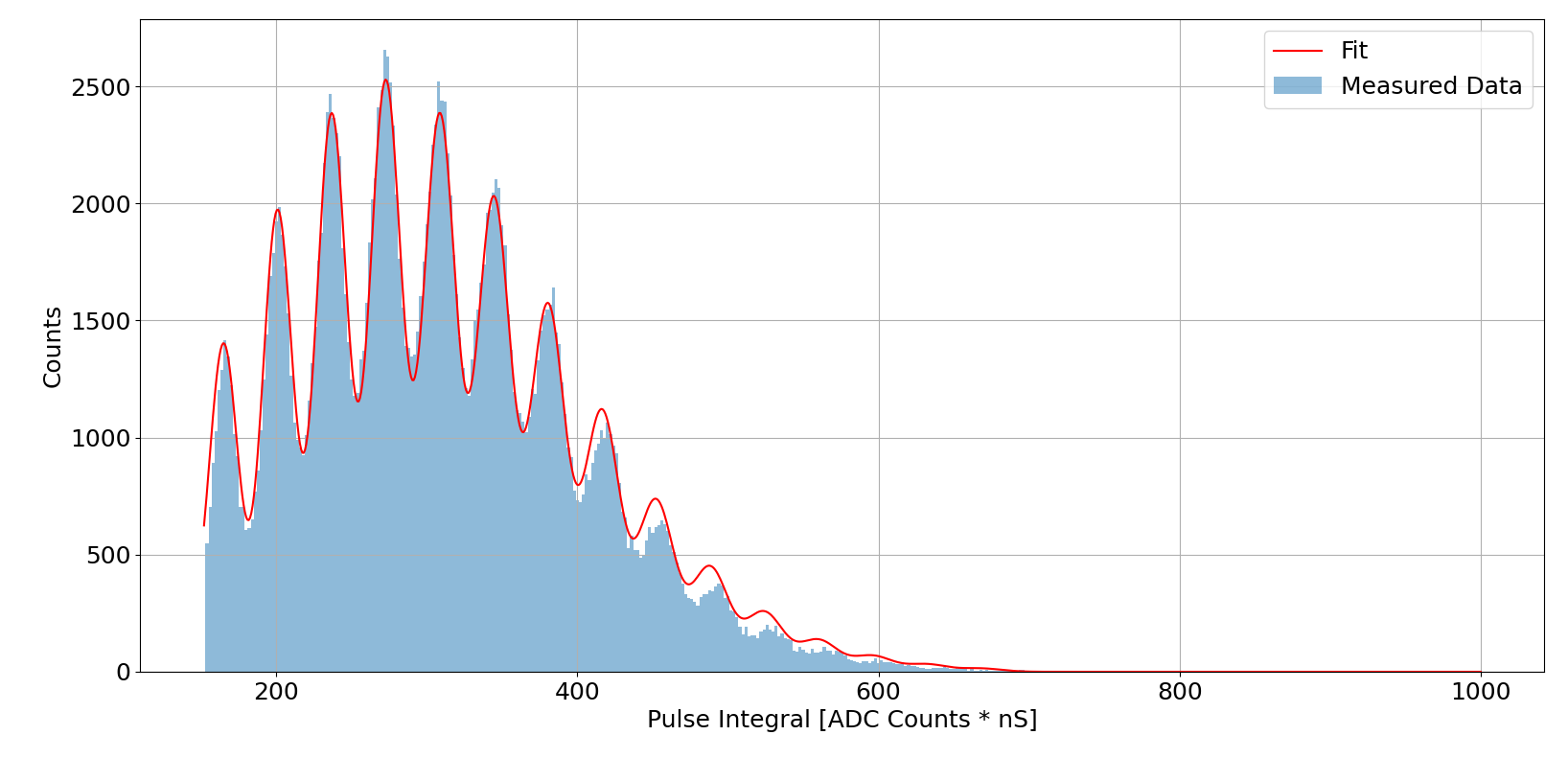}
    \caption{Left: Multi Photo-Electron (MPE) spectrum obtained with the FANSIC connected to a 5~GSps oscilloscope with 16 bits resolution, giving a single p.e. SNR of 6.5. Right: MPE obtained with the FADC board prototype with 1~GSps with 9 bits resolution, giving a single p.e. SNR of 4.5.}
    \label{fig:mpes}
\end{figure}

The AdvCam FADC board prototype developed by INFN/Padova digitizes 12 SiPM channels at 1~GS/s with 9-bit resolution. During the first stage of board development, the data was transmitted via JESD204C over Firefly optical links to the back-end board, a Kintex Ultrascale test board from Xilinx (KCU105). However, as said above, the final solution, a custom low-resource RoCEv2 RDMA core on FPGA has been deployed allowing direct data transfer to server or GPU memory. A report on the performance of this board can be found here \cite{Marini2025}.
The FADC board was successfully tested with the FANSIC chip described in Section~\ref{sec:preamp}, showing a 1~p.e. SNR of 4.5 compared to 6.5 obtained with an oscilloscope. The multi photo electron spectra in both cases are shown in Figure~\ref{fig:mpes}. 
The future version, called front-end board (FEB) aims to eliminate the front-end analogue chain, integrate White Rabbit synchronization\footnote{\href{https://white-rabbit.web.cern.ch}{https://white-rabbit.web.cern.ch}}, and scale to 49 channels with full DAQ and trigger functionality on board.

\subsubsection{The L1 trigger}
\label{sec:l1trig}
The first trigger level, dubbed L1 trigger, runs in the FEB and is the results of the discrimination of the sum of the digital signals of any combination of 49 pixels, a so-called L1 trigger region, with the granularity of a flower. 
Each FEB is connected to its six direct neighbour and collect the sum data from the neighbouring flowers, ensuring that there are no trigger dead spaces. Any positive L1 trigger leads to a second event validation performed by the second level trigger performed in the CTPB.
To achieve the second level trigger, a binary stream is built for each flower indicating whether the sum of its seven pixels is above a certain threshold, different from the L1 trigger threshold. 
The implementation of the L1 trigger as well as the binary stream for the L2 trigger is performed by CIEMAT/Spain. Up to now, the logic has been implemented on a test bench composed of an ADCQJ1600EVM TI board connected to  KCU105 Xilinx demonstration board. 
The firmware for the digital sum, as well of the generation of the binary stream for the flower threshold running at 1~GHz, have been implemented together with rate counters.
The implementation for 49 pixels and 6~$\mu$s of buffer depth fits well within the allocated budget, mostly constrained by the management of the FADC stream. 


\subsubsection{The Central Trigger Processor board}

The CTPB is being designed by the University Complutense Madrid (UCM) group to implement an L2 stereoscopic trigger combining the binary streams collected from the 163 front-end boards via optical links. 
The architecture involves three Kintex UltraScale FPGAs for processing of the trigger stream. Each of this three FPGA received data from a third of the camera pixels. Finally, a fourth FPGA for L2 logic, timing, and stereo coordination. 
Currently, two test benches are being developed: one for testing gigabit transceivers for high-performance applications and FireFly integration, and another for firmware development and testing trigger algorithms. 

\subsubsection{The L2 trigger}

The L2 trigger will start with a spatio-temporal clustering trigger specifically running prior to the stereoscopy in order to better extract the shower timing and its location in the camera FoV, allowing for more accurate coincidences: the better shower front timing extraction allows to reduce the coincidence width improving the signal to noise ratio while the shower location enable a topological trigger.
The UniGe has proposed the DBScan\footnote{Density-Based Spatial Clustering of Applications with Noise} algorithm to perform the task proving to be extremely efficient. However, porting it to FPGA is not ideal as it is not parallelisable nor has a fixed latency. The Haute École du Paysage, d'Ingénierie et d'Architecture (HEPIA) in Geneva proposes a similar approach, relying on 3D convolutions that is fully parallelisable and therefore achieves throughput of 350~MHz with latency of below 15~ns when implemented on FPGA. 
Alternatively, UCM is studying implementation of deep learning models for real-time inference. Based on the CTLearn~\cite{Brill2018}, light models are being tested to perform discriminations between noise events and showers, as well as between gamma and hadrons. Until now, throughput of few tens of kHz have been achieved with latencies of the order of few $\mu$s. At this moment, these performance makes it more suitable for running after stereoscopy.


More details about those algorithms and the trigger gain are given in~\cite{Burmistrov2025}.

\subsection{Data Acquisition}
To address the high data throughput requirements of the AdvCam, the RoCE protocol has been adopted for its low-latency, high-bandwidth capabilities and compatibility with off-the-shelf hardware, such as NVIDIA ConnectX network cards. This allows direct memory access with minimal CPU involvement, enabling real-time data transfer and processing. 
A prototype implementation by INFN/Padova demonstrated stable 9.7~Gb/s transfers using a custom RDMA core on a Xilinx FPGA, integrated within the SLAC Ultimate RTL Framework (SURF) firmware framework. The approach aligns with proven network-based event building strategies in particle physics, as used at the Large Hadron Collider, and offers scalability for future developments within modern high-performance computing environments.
Once the data packets are collected, the event building will be performed allowing for further real-time deep-learning inference for better gamma/hadron rejection. 

\section{Optical module prototype}
\label{sec:optmodproto}

\begin{figure}
    \centering
    \includegraphics[width=0.7\linewidth]{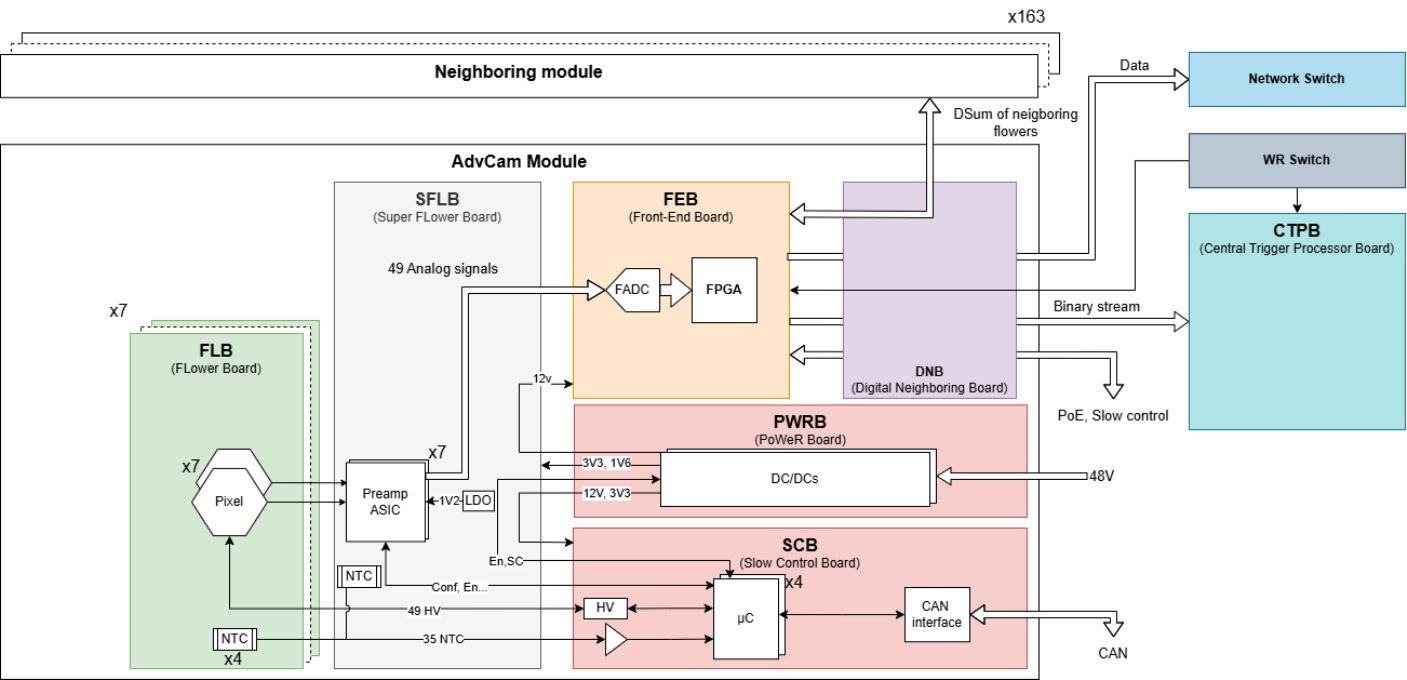} 
    \caption{Architecture of the prototype module being developed for the AdvCam.}
    \label{fig:optmod}
\end{figure}

The optical module development for AdvCam is progressing with a new architecture centered on the integration of SiPMs and with the updated FANSIC v2 ASIC on the Super Flower Board (SFLB) (see Figure~\ref{fig:optmod}).
The effort is lead by the UniGe in collaboration with INFN/Padova and Swiss industries.
The FLB schematic is under outsourced design, with interface pinouts defined for connections to both the Super Flower Board (SFLB).
The SFLB has finalized its architecture and interface specifications, and will include a 4-channel DAC for voltage reference control. 
Concurrently, the Slow Control Board (SCB) and Power Board (PWRB) have completed architectural definition, with active component selection underway. The next steps focus on completing schematics and layouts, and finalizing inter-board interfaces for prototype fabrication. The completion of the first module prototype is foreseen for Q1 2026.

\section{The AdvCam Monte Carlo performance}

\subsection{Stereoscopic performance using Random Forest based analysis}
We compared the instrument response Functions for LSTs equipped with the current camera versus the AdvCam, using the magic-ctapipe framework \cite{Abe2023b}, which allows for stereoscopic reconstruction.
The simulation results show that The AdvCam configuration significantly enhances sensitivity below 40~GeV compared to the current PMT cameras with standard analysis cuts. It is worth mentioning that the improvement is not only related to the replacement of the PMTs by SiPMs, which increase the sensitivity to Cherenkov light, but as well by the higher image granularity.


\subsection{Stereoscopic performance using deep learning analysis}

To fully exploit the enhanced sensitivity of the AdvCam for faint low-energy events, a novel IACT analysis scheme has been developed within CTLearn~\cite{Brill2018}. 
The approach leverages advanced image cleaning using DBSCAN clustering on waveform data, generating temporally resolved masks that preserve shower development. 
These masks feed into convolutional neural networks (CNNs) for gamma/hadron separation, energy, and direction reconstruction. 
The unprecedented challenge is to cope with the abundance of faint and nearly featureless images that are recovered. The preliminary results are promissing but the training process requires additional development to fully exploit those faint events.



\section{Conclusions}
The architecture of the AdvCam architecture has been defined and first prototypes of various sub-systems have been built for validation. The required performance being met, the collaboration is now focusing on interfacing all the sub-systems together and scaling them to the final number of channels. A first complete optical module is expected early 2026.
Recent Monte-Carlo simulations and analysis with state of the art data pipelines confirm the improved performance of the AdvCam, especially below 40~GeV. 
However, lowering the energy threshold, thanks to the sensor sensitivity increase and enhanced triggering capabilities poses new challenges with the need to properly account for the abundance of faint and nearly feature less events.
It is worth eventually noting that the proposed design is perfectly suited for the Medium-Sized Telescopes of CTAO as well.


\newpage

\textbf{Full Author List: CTAO-LST Project}

\tiny{\noindent
K.~Abe$^{1}$,
S.~Abe$^{2}$,
A.~Abhishek$^{3}$,
F.~Acero$^{4,5}$,
A.~Aguasca-Cabot$^{6}$,
I.~Agudo$^{7}$,
C.~Alispach$^{8}$,
D.~Ambrosino$^{9}$,
F.~Ambrosino$^{10}$,
L.~A.~Antonelli$^{10}$
% ...
The CTAO-LST Project
}\\

\tiny{\noindent$^{1}${Department of Physics, Tokai University, 4-1-1, Kita-Kaname, Hiratsuka, Kanagawa 259-1292, Japan}.
$^{2}${Institute for Cosmic Ray Research, University of Tokyo, 5-1-5, Kashiwa-no-ha, Kashiwa, Chiba 277-8582, Japan}.
$^{3}${INFN and Università degli Studi di Siena, Dipartimento di Scienze Fisiche, della Terra e dell'Ambiente (DSFTA), Sezione di Fisica, Via Roma 56, 53100 Siena, Italy}.
$^{4}${Université Paris-Saclay, Université Paris Cité, CEA, CNRS, AIM, F-91191 Gif-sur-Yvette Cedex, France}.
$^{5}${FSLAC IRL 2009, CNRS/IAC, La Laguna, Tenerife, Spain}.
$^{6}${Departament de Física Quàntica i Astrofísica, Institut de Ciències del Cosmos, Universitat de Barcelona, IEEC-UB, Martí i Franquès, 1, 08028, Barcelona, Spain}.
$^{7}${Instituto de Astrofísica de Andalucía-CSIC, Glorieta de la Astronomía s/n, 18008, Granada, Spain}.
$^{8}${Department of Astronomy, University of Geneva, Chemin d'Ecogia 16, CH-1290 Versoix, Switzerland}.
$^{9}${INFN Sezione di Napoli, Via Cintia, ed. G, 80126 Napoli, Italy}.
$^{10}${INAF - Osservatorio Astronomico di Roma, Via di Frascati 33, 00040, Monteporzio Catone, Italy}.
% ...
}

\acknowledgments 
\tiny{
We gratefully acknowledge financial support from the following agencies and organisations:
Conselho Nacional de Desenvolvimento Cient\'{\i}fico e Tecnol\'{o}gico (CNPq), Funda\c{c}\~{a}o de Amparo \`{a} Pesquisa do Estado do Rio de Janeiro (FAPERJ), Funda\c{c}\~{a}o de Amparo \`{a} Pesquisa do Estado de S\~{a}o Paulo (FAPESP), Funda\c{c}\~{a}o de Apoio \`{a} Ci\^encia, Tecnologia e Inova\c{c}\~{a}o do Paran\'a - Funda\c{c}\~{a}o Arauc\'aria, Ministry of Science, Technology, Innovations and Communications (MCTIC), Brasil;
Ministry of Education and Science, National RI Roadmap Project DO1-153/28.08.2018, Bulgaria;
Croatian Science Foundation (HrZZ) Project IP-2022-10-4595, Rudjer Boskovic Institute, University of Osijek, University of Rijeka, University of Split, Faculty of Electrical Engineering, Mechanical Engineering and Naval Architecture, University of Zagreb, Faculty of Electrical Engineering and Computing, Croatia;
Ministry of Education, Youth and Sports, MEYS  LM2023047, EU/MEYS CZ.02.1.01/0.0/0.0/16\_013/0001403, CZ.02.1.01/0.0/0.0/18\_046/0016007, CZ.02.1.01/0.0/0.0/16\_019/0000754, CZ.02.01.01/00/22\_008/0004632 and CZ.02.01.01/00/23\_015/0008197 Czech Republic;
CNRS-IN2P3, the French Programme d’investissements d’avenir and the Enigmass Labex, 
This work has been done thanks to the facilities offered by the Univ. Savoie Mont Blanc - CNRS/IN2P3 MUST computing center, France;
Max Planck Society, German Bundesministerium f{\"u}r Bildung und Forschung (Verbundforschung / ErUM), Deutsche Forschungsgemeinschaft (SFBs 876 and 1491), Germany;
Istituto Nazionale di Astrofisica (INAF), Istituto Nazionale di Fisica Nucleare (INFN), Italian Ministry for University and Research (MUR), and the financial support from the European Union -- Next Generation EU under the project IR0000012 - CTA+ (CUP C53C22000430006), announcement N.3264 on 28/12/2021: ``Rafforzamento e creazione di IR nell’ambito del Piano Nazionale di Ripresa e Resilienza (PNRR)'';
ICRR, University of Tokyo, JSPS, MEXT, Japan;
JST SPRING - JPMJSP2108;
Narodowe Centrum Nauki, grant number 2023/50/A/ST9/00254, Poland;
The Spanish groups acknowledge the Spanish Ministry of Science and Innovation and the Spanish Research State Agency (AEI) through the government budget lines
PGE2022/28.06.000X.711.04,
28.06.000X.411.01 and 28.06.000X.711.04 of PGE 2023, 2024 and 2025,
and grants PID2019-104114RB-C31,  PID2019-107847RB-C44, PID2019-104114RB-C32, PID2019-105510GB-C31, PID2019-104114RB-C33, PID2019-107847RB-C43, PID2019-107847RB-C42, PID2019-107988GB-C22, PID2021-124581OB-I00, PID2021-125331NB-I00, PID2022-136828NB-C41, PID2022-137810NB-C22, PID2022-138172NB-C41, PID2022-138172NB-C42, PID2022-138172NB-C43, PID2022-139117NB-C41, PID2022-139117NB-C42, PID2022-139117NB-C43, PID2022-139117NB-C44, PID2022-136828NB-C42, PDC2023-145839-I00 funded by the Spanish MCIN/AEI/10.13039/501100011033 and “and by ERDF/EU and NextGenerationEU PRTR; the "Centro de Excelencia Severo Ochoa" program through grants no. CEX2019-000920-S, CEX2020-001007-S, CEX2021-001131-S; the "Unidad de Excelencia Mar\'ia de Maeztu" program through grants no. CEX2019-000918-M, CEX2020-001058-M; the "Ram\'on y Cajal" program through grants RYC2021-032991-I  funded by MICIN/AEI/10.13039/501100011033 and the European Union “NextGenerationEU”/PRTR and RYC2020-028639-I; the "Juan de la Cierva-Incorporaci\'on" program through grant no. IJC2019-040315-I and "Juan de la Cierva-formaci\'on"' through grant JDC2022-049705-I. They also acknowledge the "Atracci\'on de Talento" program of Comunidad de Madrid through grant no. 2019-T2/TIC-12900; the project "Tecnolog\'ias avanzadas para la exploraci\'on del universo y sus componentes" (PR47/21 TAU), funded by Comunidad de Madrid, by the Recovery, Transformation and Resilience Plan from the Spanish State, and by NextGenerationEU from the European Union through the Recovery and Resilience Facility; “MAD4SPACE: Desarrollo de tecnolog\'ias habilitadoras para estudios del espacio en la Comunidad de Madrid" (TEC-2024/TEC-182) project funded by Comunidad de Madrid; the La Caixa Banking Foundation, grant no. LCF/BQ/PI21/11830030; Junta de Andaluc\'ia under Plan Complementario de I+D+I (Ref. AST22\_0001) and Plan Andaluz de Investigaci\'on, Desarrollo e Innovaci\'on as research group FQM-322; Project ref. AST22\_00001\_9 with funding from NextGenerationEU funds; the “Ministerio de Ciencia, Innovaci\'on y Universidades”  and its “Plan de Recuperaci\'on, Transformaci\'on y Resiliencia”; “Consejer\'ia de Universidad, Investigaci\'on e Innovaci\'on” of the regional government of Andaluc\'ia and “Consejo Superior de Investigaciones Cient\'ificas”, Grant CNS2023-144504 funded by MICIU/AEI/10.13039/501100011033 and by the European Union NextGenerationEU/PRTR,  the European Union's Recovery and Resilience Facility-Next Generation, in the framework of the General Invitation of the Spanish Government’s public business entity Red.es to participate in talent attraction and retention programmes within Investment 4 of Component 19 of the Recovery, Transformation and Resilience Plan; Junta de Andaluc\'{\i}a under Plan Complementario de I+D+I (Ref. AST22\_00001), Plan Andaluz de Investigaci\'on, Desarrollo e Innovación (Ref. FQM-322). ``Programa Operativo de Crecimiento Inteligente" FEDER 2014-2020 (Ref.~ESFRI-2017-IAC-12), Ministerio de Ciencia e Innovaci\'on, 15\% co-financed by Consejer\'ia de Econom\'ia, Industria, Comercio y Conocimiento del Gobierno de Canarias; the "CERCA" program and the grants 2021SGR00426 and 2021SGR00679, all funded by the Generalitat de Catalunya; and the European Union's NextGenerationEU (PRTR-C17.I1). This research used the computing and storage resources provided by the Port d’Informaci\'o Cient\'ifica (PIC) data center.
State Secretariat for Education, Research and Innovation (SERI) and Swiss National Science Foundation (SNSF), Switzerland;
The research leading to these results has received funding from the European Union's Seventh Framework Programme (FP7/2007-2013) under grant agreements No~262053 and No~317446;
This project is receiving funding from the European Union's Horizon 2020 research and innovation programs under agreement No~676134;
ESCAPE - The European Science Cluster of Astronomy \& Particle Physics ESFRI Research Infrastructures has received funding from the European Union’s Horizon 2020 research and innovation programme under Grant Agreement no. 824064.}

\end{document}